\documentclass[aps,reprint,groupedaddress,nofootinbib,amsmath,amssymb,floatfix]{revtex4-2}

\usepackage{graphicx}
\usepackage{dcolumn}
\usepackage{bm}
\usepackage{hyperref}
\hypersetup{
     colorlinks   = true,
     citecolor    = blue
}
\everymath{\displaystyle}
\usepackage{pifont}
\usepackage{comment}
\usepackage{bm}
\usepackage{url}
\usepackage{braket}

\usepackage{ytableau}
\everymath{\displaystyle}
\usepackage{cancel}
\usepackage{mathtools}
\usepackage{soul}

\newcommand{\be}{\begin{equation}}
\newcommand{\ee}{\end{equation}}
\newcommand{\ba}{\begin{eqnarray}}
\newcommand{\ea}{\end{eqnarray}}

\everymath{\displaystyle}

\begin{document}

\begin{flushright}
CERN-TH-2022-061
\end{flushright} 
\title{Dynamically Induced Topological Inflation}

\newcommand{\TDLI}{\affiliation{Tsung-Dao Lee Institute (TDLI) \& School of Physics and Astronomy, Shanghai Jiao Tong University, \\ Shengrong Road 520, 201210 Shanghai, P.\ R.\ China}}

\author{Gongjun Choi}
\email{gongjun.choi@cern.ch}
\affiliation{Theoretical Physics Department, CERN, CH-1211 Gen\`eve 23, Switzerland}

\author{Weikang Lin}
\email{weikanglin@sjtu.edu.cn}
\TDLI

\author{Tsutomu T. Yanagida}
\email{tsutomu.tyanagida@sjtu.edu.cn}
\TDLI
\affiliation{Kavli IPMU (WPI), The University of Tokyo, Kashiwa, Chiba 277-8583, Japan}

\begin{abstract}
We propose an inflation model in which the inflationary era is driven by the strong dynamics of $Sp(2)$ gauge theory. The quark condensation in the confined phase of $Sp(2)$ gauge theory generates the inflaton potential comparable to the energy of the thermal bath at the time of phase transition. Afterwards, with super-Planckian global minimum, the inflation commences at a false vacuum region lying between true vacuum regions and hence the name ``topological inflation". Featured by the huge separation between the scale of the false vacuum ($V(0)^{1/4}\sim10^{15}{\rm GeV}$) and the global minimum ($\langle\phi\rangle\sim M_{P}$), the model can be consistent with CMB observables without suffering from the initial condition problem. Crucially, this is achieved without any fine-tuning of parameters in $V(\phi)$. In addition to $Sp(2)$, this model is based on an anomaly free $Z_{6R}$ discrete $R$ symmetry.  Remarkably, while all parameters are fixed by CMB observations, the model predicts a hierarchy of energy scales including the inflation scale, SUSY-breaking scale, R-symmetry breaking scale, Higgsino mass and the right-handed neutrino mass given in terms of the dynamical scale of $Sp(2)$.

\end{abstract}

\maketitle


{\bf \textit{Introduction}\;---} The cosmic inflation has been the firmly established solution to the main cosmological issues including the flatness problem and the horizon problem~\cite{Starobinsky:1980te,Guth:1980zm,Linde:1981mu,Albrecht:1982wi}. The requirement for the de Sitter period and its end leads on to the slow-roll inflation models~\cite{Linde:1981mu,Albrecht:1982wi}. In these models, starting from the origin in the field space, the inflaton field ($\phi$) goes through the slow-rolling and eventually arrives at the global minimum, which results in the end of inflation.  The requisite flatness near the origin, however, may require the unnatural fine-tuning of parameters of an inflaton potential. Moreover, more severe is how to justify the initial location of $\phi$ near the origin from the outset.

These problems bothering the new inflation type models, nevertheless, could be overcome provided the inflaton potential ($V(\phi)$) is featured by a noticeable hierarchy between the scale of the false vacuum ($V(0)^{1/4}$) and the global minimum ($\phi_{\rm min}$), and $\mathcal{O}(1)$ parameters in $V(\phi)$. Particularly the hierarchy concerns the initial location of the inflaton ($\phi_{\rm ini}$). In the case where $V(0)^{1/4}$ and $\langle\phi\rangle$ coincide with a scale of the phase transition (PT) in the pre-inflationary era, $\phi_{\rm ini}$ on the hill top is not guaranteed because the field fluctuation $\delta\phi\sim H\sim T^{2}$\footnote{In this paper, we invoke the Planck unit in which the reduced Planck scale is set to the unity, i.e. $M_{P}=(8\pi G)^{-1/2}=1$.} becomes comparable to $\phi_{\rm min}$. For that reason, it becomes questionable whether the inflationary era can start in the small field inflation models after PT.\footnote{By the small field inflation models, we mean the models where $\phi_{\rm ini}$ starts the slow-roll near the false vacuum of $V(\phi)$ which is taken to be located at $\phi=0$. Usually a PT from a symmetry breaking precedes the inflationary era, generating $V(\phi)$ for inflation. The ``new inflation"~\cite{Linde:1981mu,Albrecht:1982wi} and the ``natural inflation"~\cite{Freese:1990rb} are the well-known examples of the small field inflation scenario.} Aside from that, it is challenging in most cases to establish the long enough flatness for the successful inflation consistent with CMB observables without relying on the artificial tuning of parameters in $V(\phi)$.        

In this work, given the aforementioned problems of the small field inflation, we propose an inflation model which establishes the separation of scales of $V(0)^{1/4}$ and $\langle\phi\rangle$ without any fine-tuning of parameters appearing in $V(\phi)$. To this end, we consider the supersymmetric $Sp(2)$ gauge theory with $N_{F}=6$ flavors of quarks $Q_{i}$ transforming as the fundamental representation. With an integer $R$-charge of $Q_{i}$, $Z_{6R}$ can be easily shown to be free of the mixed anomaly  $Z_{6R}-[Sp(2)]^{2}$ and thus we choose $Z_{6R}$ as the R-symmetry of the theory. On top of this, because the discrete $Z_{6}$ is also free of the mixed anomaly $Z_{6}-[Sp(2)]^{2}$ within $Sp(2)$ gauge theory, we introduce $Z_{6}$ as a gauge symmetry of the theory. Once the $Sp(2)$ gauge theory enters the confinement phase at the dynamical scale $\Lambda_{*}$, the quark condensate $\langle QQ\rangle$ forms to produce $V(\phi)$~\cite{Seiberg:1994bz}. As such, the scale of $\langle QQ\rangle\sim\Lambda_{*}^{2}$ determines the inflation scale. We shall show that how the presence of $Z_{6}$ resolves the fine-tuning of parameters in $V(\phi)$ which the similar set-up without $Z_{6}$ suffers from~\cite{Izawa:1998rh,Choi:2022fce}. In this way, we trade the parameter fine-tuning problem with the additional symmetry, which is nonetheless non-trivial. 

Quark fields $Q_{i}$ being charged under $Z_{6R}$, its condensation in the non-perturbative regime of $Sp(2)$ drives the spontaneous breaking of $Z_{6R}$ to $Z_{2R}$. However, in our model, there is no domain wall problem associated with the discrete $R$-symmetry since the domain walls are diluted away by the inflation. The model gains more attractiveness when we specify the role of quark condensate more than generating the inflaton potential: as a spurion field arising from the spontaneous breaking of $Z_{6R}$, the quark condensate explains the dimensionful parameters in the MSSM. This interesting attribute of the model unifies the origin of various energy scales including inflation scale ($H_{\rm inf}$), SUSY-breaking scale ($\sqrt{F_{Z}}$), R-symmetry breaking scale, Higgsino mass ($\mu_{H}$) and the right-handed (RH) neutrino mass ($m_{N}$) and explains those based on a single energy scale $\Lambda_{*}$ inferred from the CMB observable.

{\bf \textit{Model}\;---} On top of the MSSM gauge group, we introduce
\be
 G=Sp(2)\otimes Z_{6R}\otimes Z_{6}\,.
\label{eq:symmetry}
\ee
as the additional symmetry group. The matter contents we assume are shown in Table.~\ref{table:qn}. In addition to the MSSM particle contents\footnote{$H_{u}$ and $H_{d}$ are the MSSM up-type and down-type Higgs $SU(2)_{L}$ doublets respectively. We denote other matter contents in the MSSM ($\bm{5^*}$ and $\bm{10}$) by using the representations of $SU(5)_{\rm GUT}$.} and the right-handed neutrino ($N$), $Sp(2)$ quark chiral multiplet $Q_{i}$, the singlet anti-symmetric field $S_{ij}=-S_{ji}$, the inflaton chiral multiplet $\Phi$ and the SUSY-breaking field $Z$ are newly introduced. The indices of $Q_{i}$ and $S_{ij}$ run from $1$ to $N_{F}=6$, which makes the mixed anomaly of $Z_{6R}-[Sp(2)]^{2}$ vanish with account taken of the contribution from the $Sp(2)$ gaugino~\cite{Ibanez:1991hv,Ibanez:1991pr,Ibanez:1992ji}. Thanks to this, the symmetry group $Sp(2)\otimes Z_{6R}$ is the gauged one.\footnote{One can also check that the charge assignment of the MSSM particle contents under $Z_{6R}$ makes $Z_{6R}$ anomaly free with respect to the MSSM gauge group~\cite{Evans:2011mf}.} With $Q_{i}$ assigned the same $Z_{6}$ charge as the $Z_{6R}$ charge, it is clear that $Z_{6}$ is free of anomaly with respect to $Sp(2)$ within  $Sp(2)$ sector. But since $Z_{6}$ is anomalous with respect to $SU(2)_{L}$, for arguing the gauged $Z_{6}$, we need additional fields contributing to $Z_{6R}-[SU(2)_{L}]^{2}$. We will get back to this point later.

\begin{table}[thp]
    \begin{ruledtabular}
    \begin{tabular}{rccccccccc}
            & $Q_i$ & $S_{ij}$ & $\Phi$ & $\bm{5^*}$ & $\bm{10}$ & $H_u$ & $H_d$ & $N$ & $Z$  \\
            \hline
        $Sp(2)$ & $\ytableausetup{textmode, centertableaux, boxsize=0.6em}
\begin{ytableau}
 \\
\end{ytableau}$ & - & - & - & - & - & - & - & - \\    
        $Z_{6R}$ & 1 & 0 & 3 & 0 & 0 & 2 & 2 & 0 & 4 \\
        $Z_6$ & 1 & 4 & 3 & 4 & 0 & 0 & 2 & 2 & 2  \\
    \end{tabular}
    \end{ruledtabular}
    \caption{Quantum numbers of the matter contents of the model under the additional symmetry group in Eq.~\eqref{eq:symmetry}. All the MSSM non-gauge interactions are consistent with the charge assignment.}
    \label{table:qn}
\end{table}

For a high enough energy scale where $Sp(2)$ is in its perturbative regime, the symmetry group in Eq.~\eqref{eq:symmetry} allows for the superpotential
\ba
W&\supset&-\lambda_{ij}S_{ij}Q_{i}Q_{j}+\lambda_{ij}gS_{ij}Q_{i}Q_{j}\Phi^{2}\cr\cr
&+&\lambda_{H,ijk\ell}Q_{i}Q_{j}Q_{k}Q_{\ell}H_{u}H_{d}\cr\cr
&+&\lambda_{N,ij}Q_{i}Q_{j}NN\cr\cr
&+&\lambda_{Z,ijk\ell}Q_{i}Q_{j}Q_{k}Q_{\ell}Z\,.
\label{eq:superpotential}
\ea
where all $\lambda$'s and $g$ are $\mathcal{O}(1)$ dimensionless coupling constants.\footnote{Although in principle the higher dimension operators $(c_{2n}/(2n)!)QQS(\Phi)^{2n}$ are allowed, for perturbative $c_{2n}$s they are negligible. So it suffices for us to consider the second term in LHS of Eq.~\eqref{eq:superpotential}.} The superpotential in the first line of Eq.~\eqref{eq:superpotential} is responsible for the inflationary dynamics, the second line for the Higgsino mass term (a.k.a $\mu$-term), the third line for the right-handed neutrino mass and the final line for the SUSY-breaking. As we shall see shortly, in the non-perturbative regime of $Sp(2)$, the quark condensate $\langle QQ\rangle\sim\Lambda_{*}^{2}$ and its powers generate dimensionful parameters, explaining various energy scales in the theory at the tree-level. 

As the $Sp(2)$ gauge theory becomes strongly coupled for the energy scale below $\Lambda_{*}$, it is described by 15 composite meson fields $\mathcal{M}_{ij}\equiv(4\pi)\langle Q_{i}Q_{j}\rangle/\Lambda_{*}$ with the deformed moduli constraint ${\rm Pf}(\mathcal{M}_{ij})=\Lambda_{*}^{3}$~\cite{Seiberg:1994bz}. For simplicity, we can make a choice of vacuum expectation values (VEV) of the quark fields ($Q_{i}$) such that the only nonvanishing meson fields are
\be
\langle Q_{i}Q_{i+1}\rangle=v^{2}=\frac{\Lambda_{*}^{2}}{4\pi}\quad {\rm for}\quad i=1,3,5\,,
\label{eq:vev}
\ee
where $\Lambda_{*}$ is the dynamical scale of $Sp(2)$. The quark condensation induces the spontaneous breaking of $Z_{6R}$ to $Z_{2R}$. This physics will be taken as the fundamental origin of the various energy scales in our scenario.

Given Eq.~\eqref{eq:vev}, in the non-perturbative regime of $Sp(2)$, the superpotential in Eq.~\eqref{eq:superpotential} transforms to
\ba
W_{\rm eff}&\supset&-\lambda\Lambda_{*}^{2}S(1+g\Phi^{2})\cr\cr&+&\lambda_{H}\Lambda_{*}^{4}H_{u}H_{d}+\lambda_{N}\Lambda_{*}^{2}NN+\lambda_{Z}\Lambda_{*}^{4}Z\cr\cr
&=&W_{\rm inf}+W_{H}+W_{N}+W_{\cancel{SUSY}}\,,\nonumber\\
\label{eq:superpotential2}
\ea
where all $\lambda$s without the flavor indices are $\mathcal{O}(1)$ dimensionless coupling constants obtained after re-scaling $\lambda$s in Eq.~\eqref{eq:superpotential}. Here $S$ is a linear combination of $S_{ii+1}$ with $i=1,3,5$.

{\bf \textit{Inflation {\rm (}$W_{\rm inf}${\rm )}}\;---} As mentioned above, the inflation of the universe in our scenario is attributed to the first term of Eq.~\eqref{eq:superpotential2}. Along with the K\"{a}hler potential of the form
\be
K(\Phi,S)\supset|S|^{2}+|\Phi|^{2}+c|S|^{2}|\Phi|^{2}+...\,,
\label{eq:kahler}
\ee
the $F$-term contribution of $S$ to the scalar potential of the model yields the inflaton potential below~\cite{Choi:2022fce}
\be
V(\phi)\simeq \Lambda_{*}^{4}e^{\frac{\phi^{2}}{2}}\left(1-g\frac{\phi^{2}}{2}\right)^{2}\left(1+c\frac{\phi^{2}}{2}\right)^{-1}\,,
\label{eq:VPhi}
\ee
where $\phi$ is the real part of the scalar component of $\Phi$. 

In order for the slow-roll of $\phi$ from somewhere near $\phi=0$ to the global minimum to be responsible for the early inflationary era of the universe, the power spectrum of the curvature perturbation ($P_{\zeta}(k)=A_{s} (k/k_{\star})^{n_{s}-1} $) needs to satisfy $A_{s}=2.1\times10^{-9}$, $n_{s}=0.9649\pm0.0042$ ($68\%$ C.L., Planck TT,TE,EE+lowE+lensing)~\cite{Planck:2018jri} and $r<0.036$ ($95\%$ C.L., BICEP/Keck)~\cite{BICEP:2021xfz} should hold at the CMB pivot scale $k_{\star}=0.05{\rm Mpc}^{-1}$. According to the scanning of the parameter space ($\Lambda_{*},c,g$) performed in \cite{Choi:2022fce} with $0<\phi_{\star}<1$, $\Lambda_{*}\sim10^{-3}$, $c\gtrsim0.4$, $g\sim0.3$ with $\phi_{\star}\sim0.6$ were found to provide a good fit to the CMB observables. The model predicts a rather low tensor-to-scalar ratio $r=\mathcal{O}(10^{-4})$. We note that it is remarkable for the inflation model in Eq.~\eqref{eq:VPhi} to accomplish the successful fit to CMB observables with $\mathcal{O}(1)$ dimensionless parameters, i.e. $\lambda$, $g$ and $c$. 

With the concrete inflaton potential specified above, now we discuss appealing points of our inflation model. As is clear from Eq.~\eqref{eq:VPhi}, the potential is of the typical new inflation type. Albeit similar in the shape, there is a crucial distinction of Eq.~\eqref{eq:VPhi} as compared to the original new inflation~\cite{Linde:1981mu,Albrecht:1982wi}: our inflation model is featured by the global minimum at $\phi_{\rm min}\sim2.5$ and the inflation scale $H_{\rm inf}\simeq\Lambda_{*}^{2}\sim10^{-6}<\!\!<1$. Thanks to the (super-Planckian) large enough global minimum, the inflationary expansion in a false vacuum residing in a wall (the spatial region in-between domains with $\langle\phi\rangle=\pm\phi_{\rm min}$) is guaranteed when $\phi_{\rm min}>1$ holds~\cite{Vilenkin:1994pv,Linde:1994wt}.\footnote{When the wall thickness $\delta\sim\phi_{\rm min}V(0)^{-1/2}$ is greater than $H^{-1}\sim V(0)^{-1/2}$, the false vacuum region located in the wall experiences the inflationary expansion~\cite{Vilenkin:1994pv,Linde:1994wt}. The condition $\delta>H^{-1}$ is converted to $\phi_{\rm min}>1$.} Hence, our model does not suffer from the initial condition problem. What's remarkable is that the separation of the scales of $V^{1/4}$ and $\phi_{\rm min}$, which is required by consistency with CMB observables and the use of logics in the topological inflation, is achieved without any fine-tuning of parameters in $V(\phi)$.

{\bf \textit{Pre-Inflation Era}\;---} At the Planck time $t\sim 1/M_{p}^{-1}$, there might be particle creation in an expanding background~ \cite{Parker:2012at,PhysRevLett.21.562,Zeldovich:1971mw}. Afterwards, when $Sp(2)$ gauge theory becomes strong enough, the thermal bath made of multiplets of $Q_{i}$, $S$ and $Sp(2)$ gluons is expected to form. With a certain distribution of $T$ of thermal baths in mind, we may consider for simplicity two classes of horizons with $T<\Lambda_{*}$ and $T>\Lambda_{*}$. For the former, $V(\phi)$ in Eq.~(\ref{eq:VPhi}) applies since the horizon is already in the confined phase of $Sp(2)$. In contrast, for the later case, starting from zero potential, $V(\phi)$ eventually develops to the form in Eq.~(\ref{eq:VPhi}) at $t\sim(M_{P}/\Lambda_{*})\Lambda_{*}^{-1}$. 

Once the inflaton potential is described by Eq.~(\ref{eq:VPhi}) everywhere, we can expect that there arise a pair of neighboring Hubble patches with $\langle\phi\rangle\sim\pm\phi_{\rm min}$ after a bit of time for homogenizing $\phi$ within horizons. Then inflation is initiated at $\phi=0$ residing in the wall between the Hubble patches (the space with $\phi$ field variation by $\sim2\phi_{\rm min}$)~\cite{Vilenkin:1994pv,Linde:1994wt}. Hence our model serves as a UV model for topological inflation scenario, providing the concrete picture for development of $V(\phi)$ at the pre-inflation era in accordance with the $Sp(2)$ strong dynamics.

{\bf \textit{SUSY Breaking {\rm (}$W_{\cancel{SUSY}}${\rm )}}\;---} The last term in Eq.~\eqref{eq:superpotential2} explains the (F-term) SUSY breaking. Now that the Polonyi field $Z$ is charged under both $Z_{6R}$ and $Z_{6}$, there cannot be a marginal or relevant operator purely composed of $Z$. This fact makes the last operator in Eq.~(\ref{eq:superpotential}), i.e. $\mathcal{O}\sim QQQQZ$, most relevant in the $F$-term contribution of $Z$ to the scalar potential\footnote{We introduce a pair of massive chiral multiplets $X$ and ${\bar X}$ which have a superpotential $W= ZX^2 + MX{\bar X} $. One-loop diagrams of the $X$ and ${\bar X}$ give a large positive soft mass squared for $Z$ to stabilize the potential of $Z$ at the origin \cite{Harigaya:2013ns}.} and thus SUSY-breaking is indeed safely guaranteed once $Sp(2)$ enters the confined phase.\footnote{When either of marginal or relevant operator purely made up of $Z$ is allowed in the superpotential, the scalar potential $|F_{Z}|^{2}$ can possibly have a SUSY-preserving global minimum. Then SUSY remains unbroken. Were it not for $Z_{6}$, for instance, the operator $m_{Z}Z^{2}$ with a dimensionful parameter $m_{Z}$ is allowed. Therefore, having other discrete symmetry than the discrete $R$-symmetry is the crucial point in constructing the SUSY-breaking sector in the model. One may be concerned about the unwanted operator $\sim ZH_uH_d N$. This one is not a problem since there is the parity as the remnant of $U(1)_{B-L}$ to suppress $\sim ZH_uH_d N$ to stabilize the SUSY vacuum. Under the parity, only $\bf 5^*, 10$ and $N$ transform as odd.} Even if the Polonyi field has no anti-symmetric flavor indices of $Sp(2)$ quarks from the beginning, we see that this way of SUSY-breaking is similar to IYIT dynamical SUSY-breaking~\cite{Izawa:1996pk,Intriligator:1996pu} in that $F_{Z}\neq0$ is attributable to quark condensation with the deformed moduli constraint. Given $F_{Z}\sim\Lambda_{*}^{4}$, the model's prediction for the gravitino mass becomes $m_{3/2}=F_{Z}/(\sqrt{3}M_{P})\sim\Lambda_{*}^{4}=\mathcal{O}(10^{-12})$.

{\bf \textit{Higgsino {\rm (}$W_{H}${\rm )} and RH Neutrino Mass {\rm (}$W_{N}${\rm )}}\;---} With $\lambda_{H},\lambda_{N}=\mathcal{O}(1)$, the second and third term in Eq.~\eqref{eq:superpotential2} can account for the Higgsino mass ($\mu_{H}$) term in the MSSM and the right-handed (RH) neutrino masses. Given $\Lambda_{*}\sim10^{-3}$ inferred from CMB observables, the model predicts $\mu_{H}=\mathcal{O}(10^{-12})$ and $m_{N}=\mathcal{O}(10^{-6})$. Below we discuss implications of the prediction.

Now that the soft masses are of the order $m_{3/2}=F_{Z}/(\sqrt{3}M_{P})$ in SUGRA~\cite{Nilles:1983ge}, we expect all of $m_{H_{u}}$, $m_{H_{d}}$ and $B$ to be $\mathcal{O}(m_{3/2})$. Hence, the prediction $\mu_{H}=\mathcal{O}(10^{-12})$ renders the electroweak symmetry breaking (EWSB) condition $(|\mu_{H}|^{2}+m_{H_{u}}^{2})(|\mu_{H}|^{2}+m_{H_{d}}^{2})\simeq(B\mu_{H})^{2}$ nicely satisfied. 

On the other hand, the heavy Majorana mass term $m_{N}NN$ and the symmetry allowed Yukawa interaction $\bm{5^*}H_{u}N$ in the superpotential can explain the tiny mass for the active neutrinos in the MSSM via the seesaw mechanism~\cite{Yanagida:1979as,*Yanagida:1979gs,GellMann:1980vs,Minkowski:1977sc}. Also we note that $m_{N}=\mathcal{O}(10^{-6})$ is sufficiently heavy enough not to exceed the maximal baryon asymmetry in the leptogenesis~\cite{Fukugita:1986hr,Buchmuller:2005eh}.

{\bf \textit{Constant Term in $W$}\;---} The constant term in the superpotential ($W_{0}$) needs to be generated so as to properly cancel the SUSY-breaking $F$-term contributions to the scalar potential of the model. To this end, we may consider $SU(3)$ pure gauge theory, i.e. without any matter fields, with which $Z_{6R}$ still remains gauge anomaly free. Then, once $SU(3)$ pure gauge theory enters its confinement at the scale $\Lambda'$, there arises the gaugino condensation \cite{Nilles:1982my,Veneziano:1982ah}, i.e. $\langle\lambda^{a}\lambda^{a}\rangle=32\pi^{2}\Lambda'^{3}$. This results in the effective superpotential which take responsible for the constant term in the superpotential, i.e.
\be
W_{0}=3\Lambda'^{3}=m_{3/2}\quad\Rightarrow\quad\Lambda'=\mathcal{O}(10^{-4})\,.
\label{eq:W0}
\ee

Given that $\Lambda'$ is one order of magnitude smaller than $\Lambda_{*}$, we expect that $W_{0}$ is generated after PT of $Sp(2)$ and during inflation. $\langle\lambda^{a}\lambda^{a}\rangle$ respecting $Z_{2R}$ but not $Z_{6R}$, there could be formation of the domain wall due to the gaugino condensation. This wall, however, will be diluted away during inflation

{\bf \textit{Cosmological Constant}\;---}
Because the model is embedded in the SUGRA framework, the vanishingly small vacuum energy (cosmological constant) remains to be discussed given the scales for the SUSY-breaking $F_{Z}$ and the R-symmetry breaking $m_{3/2}$. In a SUGRA model, the difference between $|F_{Z}|^{2}$ and $m_{3/2}^{2}$ determines the scalar potential at the leading order. In our model, since both of $F_{Z}$ and $m_{3/2}$ are of the order $\mathcal{O}(\Lambda_{*}^{4})$, the required cancellation among the two can be indeed well achieved. Behind this cancellation is the $Sp(2)$ strong dynamics as the common origin of operators $QQQQH_{u}H_{d}$ and $QQQQZ$. Because $\mu_{H}$-parameter along with the EWSB was decisive for determining the scale of $m_{3/2}$, the quantum number of $H_{u}H_{d}$ was referred in determining that of $Z$.

{\bf \textit{Discussion and Outlook}\;---} In this letter, we proposed a new inflation model with the huge separation between the scale of the false vacuum ($V(\phi)^{1/4}\sim10^{15}{\rm GeV}$) and the global minimum ($\langle\phi\rangle\simeq M_{P}$). We discussed how the physics in the pre-inflationary era in our model can justify occurrence of the inflation in the similar way to topological inflation. Thereby the model was shown free from the initial condition problem. What's new as compared to the arguments made in the topological inflation scenarios~\cite{Vilenkin:1994pv,Linde:1994wt} lies in $Sp(2)$ strong dynamics-driven inflaton potential generation which importantly does not have any fine-tuned parameter.  

The interesting observation that $Z_{6R}$ and $Z_{6}$ are gauge anomaly free with respect to $Sp(2)$ within the $Sp(2)$ sector was invoked for achieving $V(\phi)$ without any tuning. The inflation scale was determined by the R-charged quark condensate and thus $R$-symmetry breaking scale, $\Lambda_{*}\sim10^{-3}$, could be inferred from CMB observables. As the spurion field of $Z_{6R}$ breaking, the quark condensate $\langle QQ\rangle\sim\Lambda_{*}^{2}$ also determines the Higgsino mass and the RH neutrino mass and thereby unifies the origin of various energy scales (the inflation scale, SUSY-breaking scale, R-symmetry breaking scale, Higgsino mass and RH neutrino mass).

We also emphasize the key idea that the model underlies. The specialty of $R$-symmetry to apply to every operator in a superpotential $W$ was used for explaining various energy scales and dimensionful parameters appearing in $W$. When every physics and operator yielding certain energy scales of interest is associated with a common symmetry, if experimental consistency allows, one may dream of explaining those energy scales as proper powers of spurion field of the symmetry. For our work, the symmetry was $Z_{6R}$.

With all phenomenologies handled in this work described by a single energy scale $\Lambda_{*}$, the model proposes the strong dynamics of $Sp(2)$ as the fundamental underlying physics governing the universe. Albeit ambitious, there are still open questions to be answered. $Z_{6}$ in Eq.~\eqref{eq:symmetry} still remains anomalous for $SU(2)_{L}$ and thus poses the question about extension of $SU(2)_{L}$-charged particle contents\footnote{One way to eliminate such an anomaly is to introduce $H'_d$ and $H'_u$ whose product $H'_uH'_d$ has charges $(2,4)$ for $Z_{6R}$ and $Z_{6}$ respectively.}. On the other hand, as the SUGRA inflation model, the model is still subject to the long-standing $\eta$-problem, i.e. how to justify suppression of all the unwanted higher dimension operators in Eq.~\eqref{eq:kahler} contributing to $V(\phi)$. We leave these structural problems of the model as the future work.

\begin{acknowledgments}
T.T.Y. appreciates Li Fu for encouraging him to consider the origin of the universe.
T.\ T.\ Y.\ is supported in part by the China Grant for Talent Scientific Start-Up Project and by Natural Science Foundation of China (NSFC) under grant No.\ 12175134 as well as by World Premier International Research Center Initiative (WPI Initiative), MEXT, Japan.
\end{acknowledgments}

\bibliography{main}

\end{document}